\documentclass[12pt]{article}
\input{amssym.def}
\input{amssym}
\newcommand{\ba}{\begin{array}}
\newcommand{\ea}{\end{array}}
\newcommand{\beq}{\begin{equation}}
\newcommand{\eeq}{\end{equation}}
\newcommand{\bea}{\begin{eqnarray}}
\newcommand{\eea}{\end{eqnarray}}
\newcommand{\beal}{\setcounter{letter}{1} \begin{eqnarray}}
\newcommand{\eeal}{\addtocounter{equation}{1} \end{eqnarray}}
\newcommand{\none}{\nonumber \\}
\newcommand{\req}[1]{Eq.(\ref{#1})}

\newcommand{\larrow}{\,\,\,\,\hbox to 30pt{\rightarrowfill}
\,\,\,\,}
\newcommand{\slarrow}{\,\,\,\hbox to 20pt{\rightarrowfill}
\,\,\,}
\newcommand{\half}{{1\over2}}
\newcommand{\IR}{{\rm I\kern-.22em R}}
\begin{document}

\begin{titlepage}
\renewcommand{\thefootnote}{\fnsymbol{footnote}}
\renewcommand{\baselinestretch}{1.3}
\medskip
\hfill  UNB Technical Report 99-03\\[20pt]

\begin{center}
{\large {\bf Boundary Dynamics of Higher Dimensional AdS Spacetime}}
 \\ \medskip  {}
\medskip
%\vfill

\renewcommand{\baselinestretch}{1}
{\bf
J. Gegenberg $\dagger$
G. Kunstatter $\sharp$
\\}
\vspace*{0.50cm}
{\sl
$\dagger$ Dept. of Mathematics and Statistics,
University of New Brunswick\\
Fredericton, New Brunswick, Canada  E3B 5A3\\
{[e-mail: lenin@math.unb.ca]}\\ [5pt]
}
{\sl
$\sharp$ Dept. of Physics and Winnipeg Institute of
Theoretical Physics, University of Winnipeg\\
Winnipeg, Manitoba, Canada R3B 2E9\\
{[e-mail: gabor@theory.uwinnipeg.ca]}\\[5pt]
 }

\end{center}

\renewcommand{\baselinestretch}{1}

\begin{center}
{\bf Abstract}
 \end{center}
{\small
We examine the dynamics
induced on the four dimensional boundary of a five dimensional
anti-deSitter spacetime by the five dimensional Chern-Simons
theory with gauge group the direct product of $SO(4,2)$ with
$U(1)$. We show
that, given boundary conditions compatible with the geometry of
5d AdS spacetime in the asymptotic region, the induced
surface theory is the $WZW_4$ model.
}
\vfill
\hfill January, 2000
\end{titlepage}

It may be possible to understand the physics of fields which
propogate through space in terms of physics on the {\it boundary} of
space \cite{hol}.  For example, even though 3d Einstein gravity
has no local dynamics,
black holes possess thermodynamic properties.  The resolution of this
seeming paradox is that the dynamical processes responsible reside on the
boundary of 3d spacetime.  This idea was broached by Carlip \cite{carlip0}
with the boundary dynamics occurring on the event horizon.
Later Strominger \cite{strom}, motivated by earlier work of
Brown and Henneaux \cite{brown} and by string theory, showed that
states on the boundary at infinity could also account for 3d black hole
thermodynamics.  This has been generalized in a variety of ways to
different species of black holes in higher dimensions \cite{epp,carlip2,solod,
smolin}.

What is a common element of all implementations of this program to date
 is the construction of {\it integrable} field
theories on some boundary of spacetime.  For instance, in 3d, the boundary
theory is a version of the 2d WZW model \cite{witten84}.  Characteristically
for integrable field theories, the WZW theories contain an infinite number of
conserved charges, or equivalently, an infinite dimensional Kac-Moody
algebra obeyed by the currents.  The association of this algebra with
conformal symmetry leads to the characterization of this phenomenon as an
example of the AdS/CFT correspondence \cite{mald}.

The starting point for the construction of the boundary theories is usually
Einstein gravity or some generalization thereof.  However, the initial work
in 3d used the close correspondence of Einstein gravity in that dimensionality
with Chern-Simons theory.  The reason for this is the fact that in the latter
it is comparatively straightforward to distangle gauge
from dynamical degrees of freedom.

An important question to ask at this point is if
this situation is a  consequence of Einstein gravity in 2+1 dimensions, or is
it more
generic?
Banados {\it et. al}\cite{bgh} have analyzed higher dimensional Chern-Simons
theories on manifolds with boundary and found evidence that
the situation might be generic. In particular, they showed that the
algebras (in diverse spacetime dimensions) of charges on the boundary are
generalizations of the Kac-Moody algebra
 that they call $WZW_4$ algebras.  The latter
are known in the mathematical literature as {\it toroidal Lie algebras}
\cite{tor}.
They generalize the Kac-Moody algebra in the sense that while the
Kac-Moody algebra is the loop algebra with a central charge, the
$n$-toroidal Lie algebras replace the loop with an $n$-torus.

The $WZW_4$ theory (sometimes known as the Nair-Schiff or Kahler Chern-Simons
theory) is a four dimensional non-linear field theory which is a natural
generalization of the 2d WZW model. Some of its properties have been developed
in the papers of Nair, Nair and Schiff, and Losev {\it et. al.} \cite{wzw4}.
 For example it is classically and perhaps quantum mechanically integrable and
one-loop renormalizable.  It was shown by Inami {\it et. al.}\cite{inami}
 that the
associated current algebra is the 2-toroidal Lie algebra.  The same authors
have also formulated a $2n$-dimensional generalization of the $WZW_4$
theory \cite{inami2}.

Banados {\it et. al.}\cite{bgh} do not explicitly derive the $WZW_4$ boundary
theory
from 5d Chern-Simons. They construct the algebra of gauge charges on the
boundary under the assumption that the gauge group is the direct product of
some non-Abelian group $G$ and the Abelian group $U(1)$. The purpose of this
Letter is to fill in this last important step: we will show that
 with boundary conditions compatible with  the geometry of
5d AdS spacetime in the asymptotic region, the induced
surface theory is precisely the $WZW_4$ model.

 We consider the Chern-Simons theory on a five dimensional manifold with
boundary. It turns out that the best starting point is the six dimensional
functional cubic in the curvature \cite{bgh}:
\beq
S_6[\hat A]=\int_{M^6}\left\langle F(\hat A)\wedge F(\hat A)\wedge F(\hat A)
\right\rangle.\label{cubic}
\eeq
In the above the integral is over a 6-manifold $M^6$, the inner product
$<>$ is an invariant trilinear form in the algebra of some group $G$ and
which satisfies the `cyclic property' $<ABC>=<CAB>$.  Finally, $\hat A$
is the connection 1-form over $M^6$ with respect to the group $G$, with
$F(\hat A):=d\hat A +\hat A\wedge \hat A=d\hat A+\half[\hat A,\hat A]$ the
corresponding curvature.  The integral $S_6$ is a topological invariant
associated with the principal bundle with group $G$ over $M^6$.
In the following we assume that the wedge
product $\wedge $ of forms is understood and will not be explicitely written.
One can show, see e.g. \cite{bgh}, that the 6-form in the integrand above
is locally exact:  it is the exterior derivative of a 5-form $S_{CS}$
called the {\it Chern-Simons form}.
The $2n+1$ dimensional generalization of this form was added as a topological
mass term to the usual Yang-Mills lagrangian and analyzed
 by Mickelsson\cite{mick1}.

In this Letter,
we restrict to the case where the gauge group $G\cong SO(4,2)
\times U(1)$. In this we are following the lead of Banados
et. al. \cite{bgh}.\footnote{Interestingly, the group $G\cong SU(2)\times U(1)$
was considered in the mid-1980's by Mickelsson\cite{mick2} in the
context of topologically massive Yang-Mills theory.}  The reason for this is
that supergravity in
a five dimensional spacetime with a negative cosmological constant leads us
to the super anti-deSitter group, whose bosonic sector is precicely $SO(4,2)
\times U(1)$.  Moreover we consider the invariant trilinear form to be
given by:
\bea
\left\langle J_a J_b J_c\right\rangle&=&Tr(J_a J_b J_c);\none
\left\langle J_a J_b J_1\right\rangle=\left\langle J_b J_a J_1\right\rangle
&=& Tr(J_a J_b);\none
\left\langle J_a J_1 J_1\right\rangle=\left\langle J_1 J_1 J_1\right\rangle
&=&0.
\eea
In the above, the $J_a$ are the generators of $SO(4,2)$, while
$J_1$ is the generator of $U(1)$.
The connection 1-form decomposes as $\hat A=A+a=A^aJ_a+a^1J_1$.

Consider the following functional of $A,a$ on a 5-manifold $M^5$:
\beq
S_{CS}[A,a]= S_{CS}[A] + 3\int_{M^5}Tr(AdA + {2\over3}A^3) \omega,
\label{eq: decomposed action}
\eeq
where
\beq
S_{CS}[A]:=\int_{M^5} Tr\left\{A(dA)^2+ {3\over 5}
A^5+{3\over2}
A^3dA\right\}\,\,\, .
\label{eq: CS action}
\eeq
with $\omega=da$ the $U(1)$ field strength.
It is straightforward to show that the exterior derivative of the
integrand in \req{eq: decomposed action} is $<(F(\hat A))^3>$, and
hence that $S_{CS}[A,a]$ is the Chern-Simons form for $SO(4,2)\times
U(1)$ for the postulated trilinear form.

Variation of the action
\req{eq: decomposed action} yields:
\beq
\delta S_{CS}[A,a] = 3\int_{M^5} Tr(F^2 \delta A) + 6\int Tr(F\delta A)\omega
   + 3\int Tr(F^2)\delta a
  +\delta S_B,
\label{eq: variation}
\eeq
where $F=dA+A^2$ and the last term is a boundary variation:
\beq
\delta S_B =  -\oint_{\partial M^5}\left[Tr(AdA +dAA+
{3\over2}A^3+3A\omega)\delta A
     + 3 (AdA+{2\over 3} A^3)\delta a\right].
\label{eq: sb}
\eeq
If $M^5$ has a non-trivial boundary $M^4:=\partial M^5$, then we
refer to fields defined on $M^5$ as bulk modes, and those
on $M^4$ as boundary modes.
The field equations for the bulk modes of the theory are:
\bea
Tr\left\{\left( F\wedge F +2 F\wedge \omega\right)J_a\right\} &=& 0,\\
Tr(F\wedge F)=g_{ab}F^a\wedge  F^b &=&0.
\label{eq: field equations}
\eea
Unlike the three dimensional case, the above action describes a theory that
does have dynamical degrees of freedom \cite{bgh}.
However
there is a `generic' sector of the phase space wherein
there
are no dynamical degrees of freedom in the bulk of the manifold, although there
are propagating modes on the boundary \cite{bgh}.
This sector is described by solutions of the form
 $F=0$ and $\det(\omega_{\mu\nu})\neq 0$ but otherwise arbitrary.
Such solutions are generic in the sense of \cite{bgh}.
Note that the presence of the boundary variation $\delta S_B$ makes the
variational
principle of the action $S_{CS}[A,a]$ ill defined as given. It is therefore
necessary to add a compensating surface term to the action whose variation
precisely cancels $\delta S_B$. This compensating surface term is determined
by the choice of boundary conditions and plays a crucial role in determining
the boundary dynamics.

Before deriving the boundary action, we outline the proof
 that the dynamics of the physical modes in the generic
sector of
the theory does indeed take place on $M^4$.   What is required is to
 derive the symplectic form on the Hamiltonian
constraint surface and show that it is a total divergence. Following the
procedure in \cite{bgh,ban1}, one first does a standard Hamiltonian
decomposition of the fields into spatial and time components. In the
Hamiltonian context, the generic sector consists of gauge fields such that  the
spatial components of the non-Abelian field strength are zero
and the spatial part of the Abelian one form is non-degerate(i.e. invertible).
The equations of motion then guarantee that the spacetime components of the
field strength  obey the generic conditions as well.
The symplectic form is found by the varying the kinetic term in the
action under arbitrary variations
of the gauge potential in the generic sector of the  constraint surface.
Since the spatial part $\bar A$ of the  connection  is flat, we can write
$\bar A= h^{-1}\bar dh$ where $h$ is an
element
of $SO(4,2)$. In terms of $h$ we find:
\bea
\delta S_{CS} &=& \int dt\int_\Sigma \omega Tr\left[D(h^{-1}\dot{h})
D(h^{-1}\delta h)\right].\nonumber\\
    &=&-\int dt \oint_{\partial\Sigma} \omega
Tr\left[\tilde{D}(h^{-1}\dot{h})(h^{-1}\delta h)\right],
\label{eq: symp1}
\eea
where $\tilde{D}$ denotes the covariant derivative along the boundary
$\partial \Sigma$ of the spatial slice $\Sigma$.
The last line in the above was obtained by integrating by parts and using the
fact that $DD\propto \bar F=0$.
\req{eq: symp1} defines the sympletic form and shows that the physical modes
``live'' on the boundary
of $\Sigma$, in complete analogy with 3-D Chern-Simons gravity\cite{ban1}.
In terms of the Lie algebra valued charges:
\beq
Q:=J_0={1\over 2} \epsilon^{ijk} \omega_{ij} h^{-1}\partial_k h,
\eeq
 we get
\beq
\delta S_{CS} = \int dt\int_{\partial \Sigma}Tr\left(\dot{Q}  \tilde L^{-1}
\delta Q\right),
\label{eq: symp2}
\eeq
where $
\tilde L= {1\over 2} \epsilon^{ijk}\omega_{ij}\tilde{D}_k,
$
 is assumed to be invertible (i.e. we exclude zero modes). Using \req{eq:
symp2} the
Poisson bracket for these charges is :
\bea
\{Q^a(x),Q^b(y)\} &=&
{1\over2}\omega_{ij}(x)\epsilon^{ijk}\tilde{D}^{ab}_k(x)\delta^3(x-y)
\nonumber\\
&=&{1\over 2}\left(
f^a{}_{cd}g^{bc}Q^d+\epsilon^{ijk}\omega_{ij}\partial_kg^{ab}\right)
\delta^3(x-y),
\label{eq: algebra}
\eea
where the $f^a{}_{bc}$ are of the $SO(4,2)$ structure constants.
The algebra \req{eq:  algebra} is the 2-{\it toroidal Lie algebra}, that is,
the current algebra
for the $WZW_4$ theory \cite{inami}.  It is a non-trivial central extension of
the Lie algebra of SO(4,2). It  generalizes the affine Kac-Moody algebra in
that the
latter is the centrally extended loop algebra with one central charge;  while
the 2-toroidal algebra is the centrally extended algebra of maps from a 2-torus
to the (finite dimensional) Lie algebra in question. The current algebra
\req{eq: algebra} agrees with the one found by Mickelsson in the mid-1980's
in his analysis of topologically massive Yang-Mills theories in $2n+1$
dimensions \cite{mick1,mick2}.

We will now determine the  surface term whose variation cancels
the unwanted boundary term  $\delta S_B$ in \req{eq: sb} for  boundary
conditions  consistent with an asymptotic region of 5d AdS spacetime.  In the
case of
3d Chern-Simons theory, this procedure results in the WZW model on the 2d
boundary \cite{witten84}.
We start by  constructing  the $AdS_5$ geometry following Witten
\cite{wittenhol}.
Consider the quadratic form:
\beq
f(u,v,x^i):=uv-\eta_{ij}x^ix^j, \label{quad}
\eeq
where $u,v,x^i$ with $i,j,...=1,2,3,4$ are
coordinates in $\IR^6$ with the
ultrarelativistic metric
\beq
ds_6^2=-dudv+\eta_{ij}dx^idx^j,\label{ultra}
\eeq
with $\eta_{ij}$ the Minkowski
metric with signature $(-,+,+,+)$.  Now consider the quadric given by the
solution of $f(u,v,x^i)=1$, with points with an overall sign change
identified.   In a coordinate patch with $v>0$, we can solve for $u=
(1+\eta_{ij}x^ix^j)/v$ and the metric \req{ultra} reduces,
after completing the square, to the $AdS_5$ metric
\beq
ds_5^2={1\over v^2}dv^2+\eta_{ij}(dx^i-{x^i\over v}dv)(dx^j-{x^j\over v}dv).
\label{ads}
\eeq
{}From the above, we see that in a region of large fixed $v$,
the leading order terms in the frame and spin-connection fields are
simply
$e^v=0$, $e^i=dx^i$,
$w^i{}_j=0$ and $ w^i{}_v=e^i$.
The corresponding  SO(4,2) gauge potential  defined in terms of the
geometrical fields is therefore flat:
\bea
A&=& \left[\begin{array}{cc} w^a{}_b & {e^a}\\
          -{e_b}& 0 \end{array}\right] =
 \left[\begin{array}{cccc} 0 & -dx_j
& 0 \\
     dx^i& 0 &dx^i \\
     0 & -dx_j &
0
\end{array}\right].
\label{eq: pot}
\eea

The first condition we impose on the boundary is that the connection $A$
is flat. In this
case we can simplify the boundary variation by replacing $AdA$ by $-A^3$.
Moreover, we have
verified that
$Tr(A^3\delta A)=0$ for all Lie algebra valued $A$ and $\delta A$. The boundary
variation reduces to:
\beq
\delta S_B = -3\int_{M^4}Tr\left(A\delta A\omega + (AdA+{2\over 3}
A^3)\delta a
\right) \quad .
\label{eq: sb2}
\eeq
We also assume that $\omega$ is a fixed two-form on the boundary so that the
variation
$\delta a$ of the U(1) gauge potential must be of the form $\delta a=
d\lambda$, for some
scalar $\lambda$. Integrating by parts and using the fact that $F(A)=0$ we find
that the
second term in \req{eq: sb2} vanishes.

Now we have assumed that $M^5$ admits a closed invertible 2-form
$\omega$ and hence
the restriction of $\omega$ to the boundary $M^4$ is closed.
{}From the flat $SO(4,2) $ connection, one may construct a {\it flat}
frame-field compatible with a (Lorentzian) spin-connection, which
upon
Wick rotation, provides a Euclidean structure.  The flat Euclidean geometry
enables
the construction of a complex structure on $M^4$
\cite{amp}.
Hence, using this
complex structure, we can write:
\beq
\delta S_B = -3\int_{M^4} Tr(A_i \delta
A_{\bar j}-A_{\bar j}\delta A_i)
      \omega_{\bar i j}dz^i\wedge d\bar z^j\wedge d\bar z^i\wedge dz^j \quad .
\label{delta SB}
\eeq
In the above the $z^i,i,j,...=1,2$ are complex coordinates on $M^4$. The above
expression cannot be integrated for arbitrary connection $A$ on the boundary. 
One way to proceed is  to assume further boundary conditions on the connection.
If we assume, analogously to Carlip for the 2+1 dimensional case \cite{carlip1},
that
$A_i $ is fixed on the boundary, then the second term in
\req{delta SB} vanishes identically and the first term
 can be trivially integrated.
With these boundary conditions, the
complete action, including required boundary term is:
\beq
S=S_{CS}[A] + 3\int_{M^5}Tr(AdA + {2\over3}A^3) \omega + 3\oint_{M^4}
Tr(A_idz^i \wedge A_{\bar i}d\bar z^i)\wedge
      \omega \quad .
\label{final boundary action}
\eeq
Recalling that on shell $F[A]=0$, we can parametrize the gauge potential as
$A=h^{-1}dh$. What we are left with is the gauge dependent part of the action,
namely:
\beq
S[h] = -\int_{M^5} Tr(h^{-1}dh)^3\omega+3 \int_{M^4} Tr[(h^{-1}\partial
h)
     (h^{-1}{\bar \partial}h)]\omega,
\label{eq: wzw4 action}
\eeq
This is precisely the $WZW_4$ action\cite{cham,inami}.  \footnote{The term
$(1/10)\int_{M^5}Tr(h^{-1}dh)^5$ from $S_{CS}$ in \req{eq: CS action} is the
winding number of maps from
$M^5$ to 5-spheres in the group manifold SO(4,2). This can be easily shown to
be identically zero for the group $SO(4,2)$ in five dimensions. } 

To summarize, we
 have examined the 5d Chern-Simons theory with gauge group $SO(4,2)\times
U(1)$ on a manifold with boundary.  We have looked at the sector of
phase space wherein the $SO(4,2)$ connection is flat and the $U(1)$
connection is fixed but not flat.  These conditions are obeyed by
a connection constructed from the frame-fields and spin-connection of
anti-deSitter spacetime.
We have shown that the procedure first applied by Carlip to 2+1
Chern-Simons theory can also be applied in the present case. The result
is that 
non-trivial field theory induced on the boundary, namely the $WZW_4$
theory, whose action is given in \req{eq: wzw4 action}.
This theory has the 2-toroidal Lie algebra (\req{eq: algebra}) as the algebra
of observables.

It seems likely that our analysis can be extended in a straightforward manner
to the Chern-Simons action generalized to $2n+1$ dimensions. The algebra
of boundary charges in the generalized case (the ``WZW${}_{2n}$ algebra'')
has already been constructed by Banados {\it et al}\cite{bgh}. (See also
Mickelsson\cite{mick1,mick2}).
 We are currently
in the process of checking whether the  boundary theory in this case
is the corresponding ``WZW${}_{2n}$ model'' of \cite{inami2}.

Finally, we would like to mention that there exists a somewhat different
and  more natural
way to arrive at the complete action, with boundary terms, in \req{final boundary action}.\footnote{We are grateful to S. Carlip for pointing this 
out to us.} In particular, it is not necessary to impose the strong boundary
condition that $A_i$ be fixed on the boundary. Instead, one notes that
the variation of
\req{final boundary action} yields the equations of motion for the bulk
with no boundary terms if and only if the connection on the boundary is
flat and expressable locally as $A=h^{-1}dh$ where the $h$ obeys the 
WZW${}_4$ field equations. In this view, one can think of the full action
as containing  both the 5-D Chern Simons theory for the bulk modes,
and the WZW${}_4$ theory for gauge modes on the boundary. It is important
to note in this context that the gauge modes on the boundary are completely
decoupled from the bulk modes. This alternative derivation of 
\req{final boundary action} and its consequences will be discussed in a 
future publication.

\noindent {\bf Acknowledgments}

\noindent  The authors would like to think Y. Billig, M. Paranjape
and M. Visser for helpful comments.
We also acknowledge the partial support of the Natural Sciences and
Engineering Research Council of Canada.

\end{document}